\documentclass[showpacs,amssymb,aps,twocolumn]{revtex4}
\usepackage{amsmath}
\usepackage{amstext}
\usepackage{amsopn}
\usepackage{amsfonts}
\usepackage{amssymb}
\usepackage{bbm}
\usepackage{accents}
\usepackage{empheq}
\usepackage{graphicx}
\usepackage{epsf}
\usepackage{graphics}
\usepackage[latin1]{inputenc}
\def\sc{\scriptscriptstyle}

\begin{document}

\title{Causal amplitudes in the Schwinger model at finite temperature}

\author{Ashok Das,$^{a,b}$ R. R. Francisco$^{c}$ and J. Frenkel$^{c}$\footnote{$\ $ e-mail: das@pas.rochester.edu,  jfrenkel@fma.if.usp.br}}
\affiliation{$^a$ Department of Physics and Astronomy, University of Rochester, Rochester, NY 14627-0171, USA}
\affiliation{$^b$ Saha Institute of Nuclear Physics, 1/AF Bidhannagar, Calcutta 700064, India}
\affiliation{$^{c}$ Instituto de Física, Universidade de São Paulo, 05508-090, São Paulo, SP, BRAZIL}

\begin{abstract}
We show, in the imaginary time formalism, that the temperature dependent parts of all the retarded (advanced) amplitudes vanish in the Schwinger model. We trace this behavior to the CPT invariance of the theory and give a physical interpretation of this result in terms of forward scattering amplitudes of on-shell thermal particles.
\end{abstract}


\maketitle
\bigskip

 \bigskip
 
 In two earlier papers \cite{frenkel,frenkel1} we derived the complete thermal effective action for $0+1$ dimensional QED as well as for $1+1$ dimensional Schwinger model (massless QED) in the real time (closed time path) formalism \cite{ctp,das}. The structure of the effective action in terms of the doubled thermal degrees of freedom allowed us to prove algebraically that all the temperature dependent retarded (advanced) amplitudes vanish in these theories. While the algebraic proof in the real time formalism is indeed quite simple and straightforward, it does not shed any light on the physical origin of such an interesting result.
 
 On the other hand, the imaginary time formalism \cite{matsubara,das,kapusta,lebellac,evans} is the natural framework in which retarded (advanced) amplitudes can be derived conveniently without the need for doubling the degrees of freedom. Therefore, in this formalism the physical meaning of such a result may be better understood within the context of the original quantum field theory. In this brief report we undertake such an analysis for the $1+1$ dimensional Schwinger model in the imaginary time formalism which leads to consistent physical interpretations for the vanishing of retarded (advanced) amplitudes in this model. The $0+1$ dimensional model has already been studied extensively in the past in the imaginary time formalism \cite{dunne,dasdunne} and it is known that the Ward identities require any amplitude to vanish when the external energies are nonvanishing. It follows from this that the retarded (advanced) amplitudes must vanish in this theory since there is no analytic continuation available to obtain a nontrivial amplitude. Therefore, we concentrate only on the $1+1$ dimensional Schwinger model in this work. 
 
The fermion sector of the Schwinger model \cite{schwinger} is described by the $1+1$ dimensional Lagrangian density (see \cite{frenkel,frenkel1} for our notations)
\begin{equation}
 {\cal L} = \bar{\psi} \gamma^{\mu} (i\partial_{\mu} - eA_{\mu})\psi.\label{dirac}
\end{equation}
 Since the fermions are massless, the Lagrangian density naturally decomposes as
\begin{align}
 {\cal L} & = {\cal L}_{\sc R} + {\cal L}_{\sc L}\nonumber\\
 & = \psi_{\sc R}^{\dagger} (i\partial_{+} - eA_{+})\psi_{\sc R} + \psi_{\sc L}^{\dagger} (i\partial_{-} - eA_{-})\psi_{\sc L},\label{sectors}
\end{align}
into a right handed and a left handed sector where
\begin{align}
 \psi_{\sc R} & = \frac{1}{2} (\mathbbm{1} + \gamma_{5}) \psi, \quad \psi_{\sc L} = \frac{1}{2} (\mathbbm{1} - \gamma_{5}),\nonumber\\
 x^{\pm} & = \frac{x^{0}\pm x^{1}}{2},\quad \partial_{\pm} = \partial_{0}\pm \partial_{1},\quad A_{\pm} = A_{0}\pm A_{1}.
\end{align}
At zero temperature the quantum corrections couple the two sectors through the chiral anomaly. However, at finite temperature there is no ultraviolet divergence and, therefore, there is no coupling between the two sectors so that we can study each sector independently. Parenthetically we remark that in $1+1$ dimensions, the Dirac matrices satisfy special identities \cite{adilson,silvana} such as ($A\!\!\!\slash = \gamma^{\mu}A_{\mu}$, $A_{\pm}$ are the light-cone components as defined above and $u^{\mu}_{\pm} = (1,\pm 1)$ are the light-cone vectors)
\begin{align}
& {\rm Tr}\ \left(\gamma^{\mu_{1}}A\!\!\!\slash \gamma^{\mu_{2}}B\!\!\!\slash \gamma^{\mu_{3}}C\!\!\!\slash\cdots\right) = A_{+}B_{+}C_{+}\cdots u^{\mu_{1}}_{-}u^{\mu_{2}}_{-}u^{\mu_{3}}_{-}\cdots\nonumber\\
&\qquad\qquad\qquad\qquad + A_{-}B_{-}C_{-}\cdots u^{\mu_{1}}_{+}u^{\mu_{2}}_{+}u^{\mu_{3}}_{+}\cdots,
\end{align}
so that even if we work in the complete theory, the loop integrals would naturally separate into the two sectors because of such special identities. 

\begin{figure}[ht!]
\begin{center}
\includegraphics[scale=1]{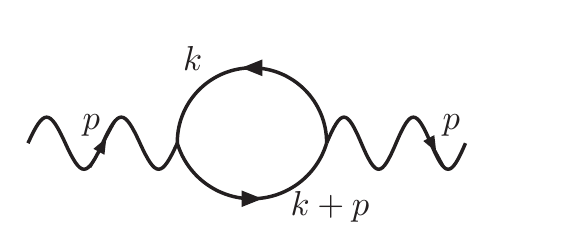}
\end{center}
\caption{Photon self-energy in the right handed sector.}
\label{f1}
\end{figure}

Let us look at the two point amplitude, Fig. \ref{f1}, in the right handed sector in the imaginary time formalism in some detail before generalizing the result to the $2n$-point amplitude (the odd point amplitudes vanish by charge conjugation symmetry).
In the imaginary time formalism, the energy becomes discrete, even multiples of $\pi T$ for bosons and odd multiples of $\pi T$ for fermions, with $T$ denoting the temperature. Correspondingly, the integral over loop energies becomes a sum over these discrete energies. As a result, the self-energy can be written as
\begin{equation}
\Pi_{\sc R} = e^{2} \int \frac{dk}{2\pi}\ T\sum_{n=-\infty}^{\infty} \frac{1}{i\omega_{n} + k} \frac{1}{i(\omega_{n}+\Omega_{\ell})+ k+p},\label{sum}
\end{equation}
where $\omega_{n} = (2n+1)\pi T, \Omega_{\ell} = 2\ell\pi T$.

\begin{figure}[ht!]
\begin{center}
\includegraphics[scale=.5]{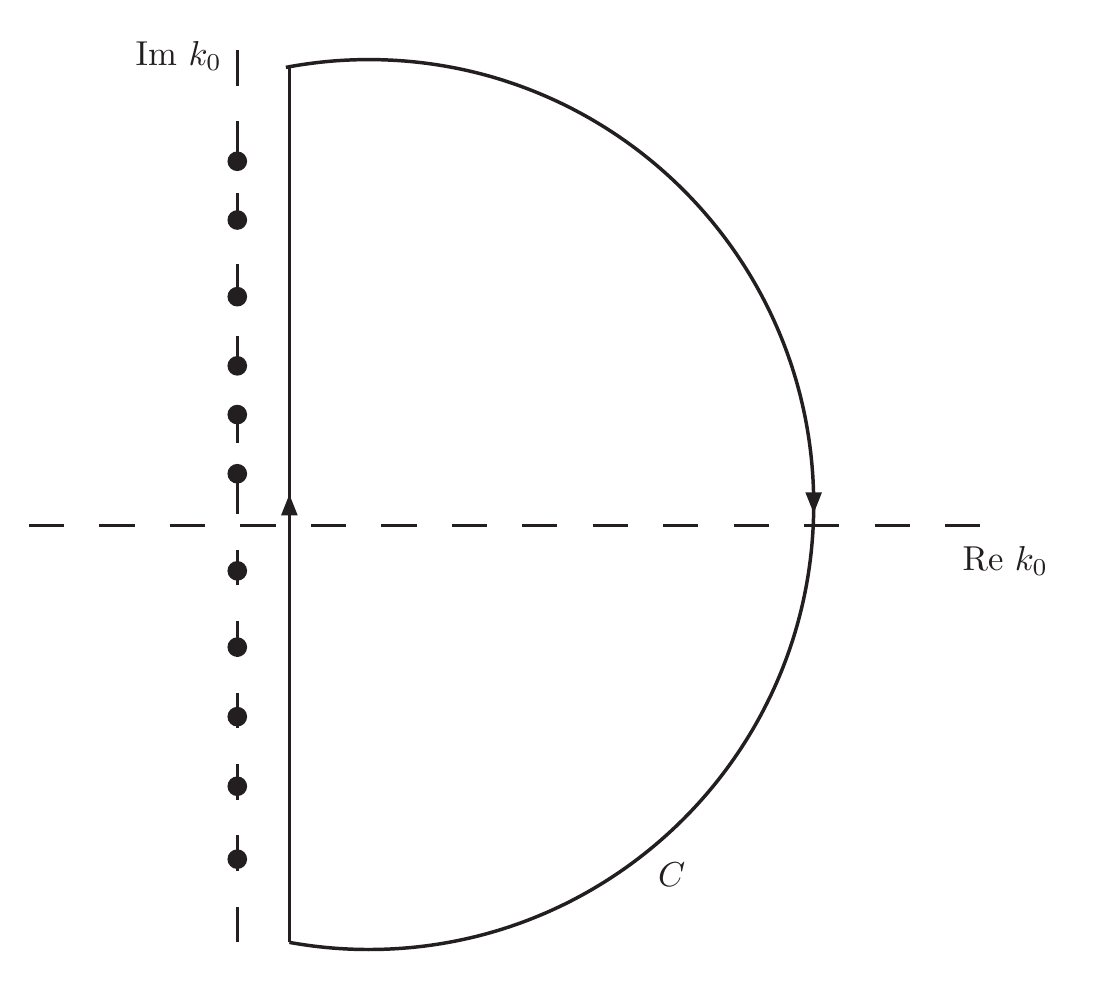}
\end{center}
\caption{Contour $C$ in the complex $k_{0}$ plane along which the integral needs to be evaluated.}
\label{f2}
\end{figure}

The sum over $n$ in \eqref{sum} can be evaluated using the method of contours \cite{kapusta,lebellac} and allows us to separate the temperature dependent part of the amplitude to be
\begin{align}
\Pi_{\sc R}^{\sc (T)} & = - \frac{e^{2}}{2\pi i} \int \frac{dk}{2\pi}\int_{\sc C} dk_{0} \Big(\frac{1}{k_{0}+k} \frac{1}{k_{0}+i\Omega_{\ell} + k+p}\nonumber\\
& \qquad\qquad + k_{0}\rightarrow -k_{0}\Big)n_{\sc F} (k_{0}),
\end{align}
where $n_{\sc F} (k_{0}) = \frac{1}{e^{k_{0}/T} + 1}$ denotes the fermion distribution function and the contour is closed (clockwise) on the positive half of the real $k_{0}$ axis (which does not enclose the poles of $n_{\sc F}(k_{0})$ along the imaginary axis) as shown in Fig. \ref{f2}. 
The contributions from the two poles in the denominators can be easily evaluated. To analytically continue to the retarded amplitude, one has to use the periodicity condition $n_{\sc F} (i\Omega_{\ell}+k+p) = n_{\sc F} (k+p)$ before letting $i\Omega_{\ell}\rightarrow p_{0}$. This leads to the temperature dependent part of the retarded self-energy to be
\begin{equation}
\Pi_{\sc R}^{\sc (T)} (p_{0},p) = -e^{2}\int \frac{dk}{2\pi}\left(\frac{\text{sgn} (k) n_{\sc F} (|k|)}{p_{0}+p} + k\rightarrow -k\right).\label{retarded}
\end{equation}
It is clear that the two terms in the integrand cancel each other leading to a vanishing temperature dependent retarded self-energy. 

\begin{figure}[ht!]
\begin{center}
\includegraphics[scale=.6]{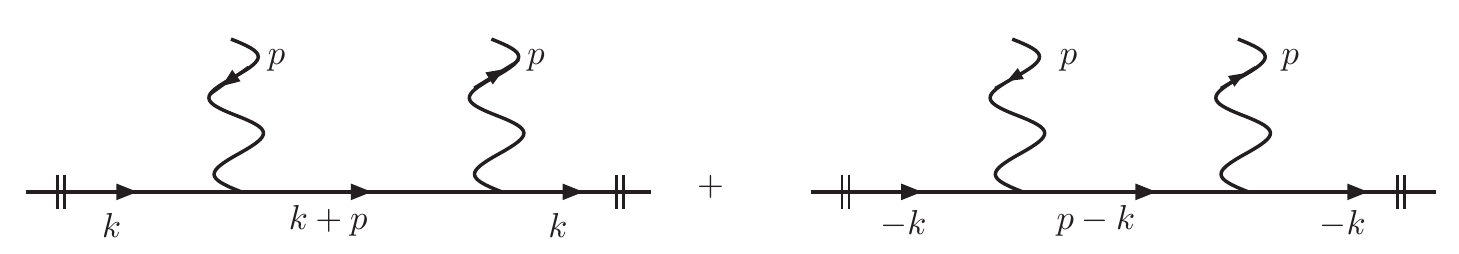}
\end{center}
\caption{Forward scattering description of the retarded self-energy. The external fermion is an on-shell thermal particle.}
\label{f3}
\end{figure}

However, to obtain a physical interpretation of this result, let us introduce an auxiliary variable of integration and rewrite \eqref{retarded} as
\begin{align}
\Pi_{\sc R}^{\sc (T)} (p_{0},p) & = e^{2}\int \frac{d^{2} k}{(2\pi)^{2}} n_{\sc F} (|k|) \Big(\frac{2\pi\delta (k_{0}+k) \text{sgn} (k_{0}-k)}{p_{0}+p+k_{0}+k}\nonumber\\
& \qquad\qquad + k\rightarrow -k\Big).\label{fsc}
\end{align}
This way of writing the self-energy makes contact with the forward scattering description of retarded amplitudes \cite{taylor,brandt1,brandt} as shown in Fig. \ref{f3}.
The two terms in the integrand correspond to the two diagrams in the forward scattering amplitude of an on-shell thermal particle. However, when the fermion is massless, changing the momentum, changes the helicity taking a particle to its anti-particle. Therefore, in this case, the two diagrams correspond to the two contributions coming from the particle and anti-particle scattering.  Cancellation between the two terms in \eqref{retarded} simply corresponds to the anti-particle contribution exactly canceling the particle contribution leading to the conclusion that in $1+1$ dimensions, an on-shell massless thermal fermion cannot scatter in the forward direction. This is not true in general, for example, if the fermion is massive this will not be the case and we will understand this result later from symmetry arguments.

\begin{figure}[ht!]
\begin{center}
\includegraphics[scale=.6]{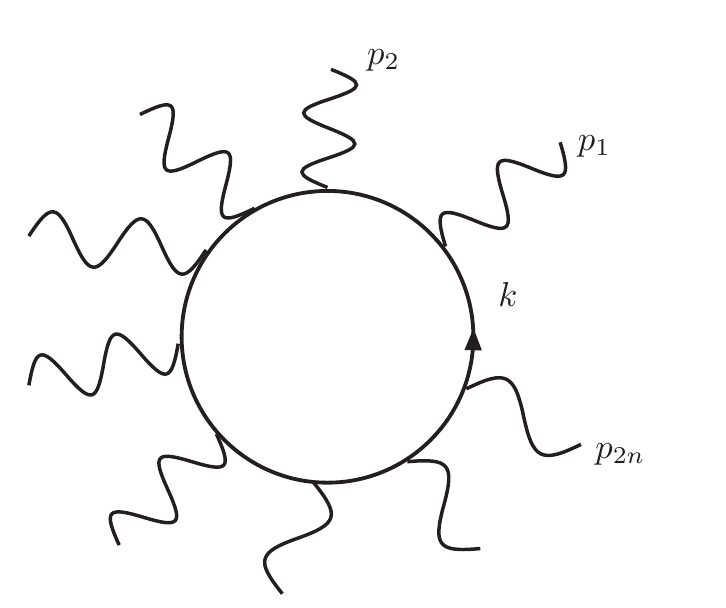}
\end{center}
\caption{$2n$-point amplitude in the right handed sector. All the external photon momenta are assumed to be incoming.}
\label{f4}
\end{figure}

A similar calculation can be carried out for the $2n$-point amplitude, Fig. \ref{f4}, in the imaginary time formalism. 
After evaluating the sum over discrete energies using the contour method and using the periodicity of the fermion distribution function in the external (bosonic) energies, we obtain the temperature dependent part of the retarded amplitude to have the form
\begin{align}
\Gamma_{2n}^{\sc (T)} & = -e^{2n}\int \frac{dk}{2\pi}\Big[ \Big(\frac{\text{sgn} (k)}{p_{1+} (p_{1+}+p_{2+})\cdots (\sum\limits_{i=1}^{2n-1} p_{i+})} + \text{perm.}\Big)\nonumber\\
& \qquad\qquad + k\rightarrow -k\Big]n_{\sc F} (|k|),\label{2npoint}
\end{align}
where the integrand again vanishes by anti-symmetry ($p_{i+}$ denote the light-cone components of the momenta). 
\begin{widetext}
\begin{center}
\begin{figure}[ht!]
\begin{center}
\includegraphics[scale=.6]{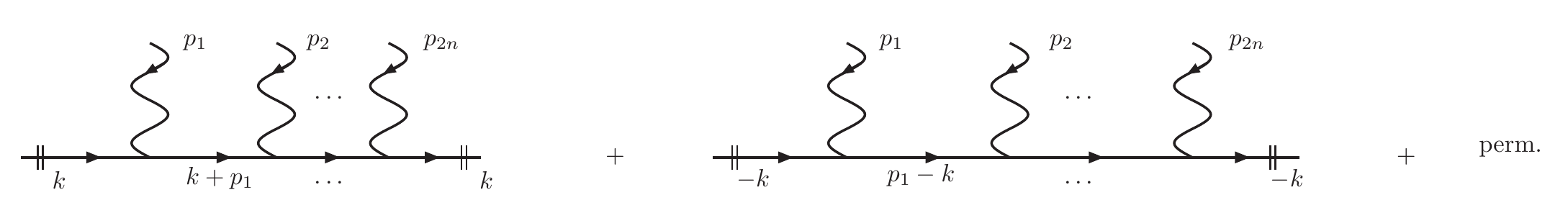}
\end{center}
\caption{Forward scattering description of the retarded $2n$-point function. The external fermion is an on-shell thermal particle and we assume $p_{1}+p_{2}+\cdots + p_{2n}=0$.}
\label{f5}
\end{figure}
\end{center}
\end{widetext}
Introducing the auxiliary variable of integration as in \eqref{fsc} we can again give this the thermal forward scattering representation as shown in Fig. \ref{f5}. The vanishing of the retarded amplitude can now be understood as arising due to the exact cancellation between the particle and the anti-particle contributions to the process.

This vanishing can also be seen directly from the basic symmetries of the theory. Let us look at the thermal effective action (in momentum space) at the $2n$-point level in the right handed sector,
\begin{align}
\Gamma^{(2n) \sc (T)} & = e^{2n}\left(\prod_{i=1}^{2n-1}\!\!\int\!\!\frac{d^{2}p_{i}}{2\pi}\right) A_{+}(p_{1})\cdots A_{+}(p_{2n-1})A_{+}(p_{2n})\nonumber\\
& \qquad \times \Gamma_{2n}^{\sc (T)} (p_{1},\cdots ,p_{2n-1}),\label{2n}
\end{align}
where $p_{2n}=-p_{1}-p_{2}-\cdots -p_{2n-1}$. CPT is an invariance of the underlying quantum field theory and, therefore, the effective action must be invariant under CPT. Under a CPT transformation,
\begin{equation}
A_{\pm} (p_{i}) \xrightarrow{CPT} -A_{\pm} (-p_{i}),
\end{equation}
so that the invariance of the effective action \eqref{2n} would require that the $2n$-point amplitude should be an even function of the external momenta,
\begin{equation}
\Gamma_{2n}^{\sc (T)} (-p_{1},\cdots, - p_{2n-1}) = \Gamma_{2n}^{\sc (T)} (p_{1},\cdots , p_{2n-1}).\label{CPT}
\end{equation}
On the other hand, we see from the explicit evaluation of the retarded amplitudes in \eqref{retarded} and \eqref{2npoint} that the dependence on the variable of integration completely drops out of each of the $(2n-1)$ denominators so that the amplitude is completely anti-symmetric
\begin{equation}
\Gamma_{2n}^{\sc (T)} (-p_{1},\cdots,-p_{2n-1}) = - \Gamma_{2n}^{\sc (T)} (p_{1},\cdots ,p_{2n-1}).
\end{equation}
Consistency with the invariance under CPT, \eqref{CPT}, then requires that the retarded amplitude must vanish and this is also consistent with the cancellation of the particle and anti-particle contributions in the forward scattering description. It is worth emphasizing that if the fermion had a mass, each of the denominators would involve a mass term and, consequently, the denominators would no longer be completely anti-symmetric and will have a symmetric part. As a result, the amplitudes will not have to vanish because of the requirement of CPT invariance. Similarly, if we are looking at a Feynman amplitude, one or more factors may involve delta functions of the external momentum which may make the amplitude symmetric so that it does not have to vanish. This anti-symmetry, therefore, is a very special feature of the retarded (advanced) amplitudes in this massless theory which is why these amplitudes vanish.

This discussion generalizes in a straightforward manner to the left handed sector (as well as to advanced amplitudes) and shows not only that the temperature dependent retarded (advanced)  amplitudes in the Schwinger model vanish, but also traces the physical reason for this to be the CPT invariance of a massless theory in $1+1$ dimensions. This also translates into an exact cancellation of the particle and anti-particle contributions in the forward scattering amplitudes.
\bigskip

\noindent{\bf Acknowledgments}
\medskip

This work was supported in part by CNPq and FAPESP (Brazil).

\end{document}